# Challenges in Analogical Reasoning

Shih-Yin Lin and Chandralekha Singh,
Department of Physics and Astronomy,
University of Pittsburgh, Pittsburgh, PA, 15260

## 1 Introduction

Learning physics is challenging. There are only a few fundamental principles of physics that are condensed into compact mathematical forms. Learning physics requires unpacking the fundamental principles and understanding their applicability in a variety of contexts that share deep features [1-7]. For example, the conservation of angular momentum principle can be used to explain why a ballerina spins faster when she puts her arms close to her body or why a neutron star collapsing under its own gravitational force spins increasingly faster as it collapses. One way to help students learn physics is via analogical reasoning [8-14]. Students can be taught to make an analogy between situations which are more familiar or easier to handle and another situation where the same physics principle is involved but that is difficult to handle. Here, we examine introductory physics students' ability to perform analogical problem solving in the context of problems involving Newton's second law. Students enrolled in an algebra-based introductory physics course were given a solved problem in a quiz involving tension in a rope and were asked to solve another problem that was very similar in the application of the physics principle but which had a different context involving friction about which students often have misleading notions. They were explicitly asked to point out the similarities between the two problems and then use the analogy to solve the quiz problem involving friction.

One common incorrect belief about the static frictional force is that it is always at its maximum value because students have difficulty with the mathematical inequality that relates the magnitude of the static frictional force to the normal force. This difficulty may partly be due to the vocabulary of introductory physics and how it is interpreted by students [15]. We find that the presence of this type of belief about friction in the quiz problem made it very difficult for the introductory students to discern the deep similarity between the two problems and exploit the analogy effectively.

## 2 The Problems used for Analogical Reasoning

The quiz problem and the solved problem in Appendix I are about a car in equilibrium on an incline. Students were asked to make an explicit analogy between the two problems and use that analogy to solve the quiz problem. Both problems are equilibrium applications of Newton's second law. Students have to realize that the car is at rest on the incline in each case, so the net force on the car is zero. Also, since the weight of the car and the normal force exerted on the car by the inclined surface are the same in both problems, the only other force acting on the car (which is the tension force in one problem and the static frictional force in the other problem) must be the same. The correct answer (the magnitudes of the tension and frictional forces) in both problems is 7,500 N, equal to the component of the weight of the car acting down the incline.

As noted earlier, introductory physics students often believe that the static frictional force must be at its maximum value $f_s^{max} = \mu_s N$ where $\mu_s$ is the coefficient of static friction and $N$ is the magnitude of the normal force. Earlier we performed a study in which some introductory physics students were asked to solve both problems involving friction and tension shown in Appendix I (no solved problem provided) and others were asked to solve only the quiz problem with friction in a multiple-choice format [16,17]. From prior experience, we knew that students in general struggle significantly more on the problem with friction than the problem with tension. We wanted to investigate if student performance was significantly different for the case in which students worked only on the problem with friction vs. when that problem was preceded by the problem involving tension in the cable. In particular, we were exploring if students who solved both problems observed the underlying similarity of the two problems and realized that the static frictional force was less than its maximum value and exactly identical to the tension in the cable in the problem pair.

The 81 students who only solved the problem with friction obtained an average of 20% whereas 479 students who solved both problems obtained 28% and 67% on the problems involving friction and tension, respectively. There is no significant difference in the performance of students who were only asked the problem with friction vs. those who were also given the problem pair with tension (despite the fact that the students performed better on the problem with tension). Thus, giving both problems did not improve student performance on the problem with friction. In other words, a majority of students who were given both problems did not discern the underlying similarity of the two problems. The two most common incorrect responses to the problem with friction were that the magnitude of the frictional force is $\mu_s N$ or $\mu_k N$. In individual discussions, students often noted that the problem with friction must be solved differently from the problem involving tension because there is a special formula for the frictional force. Even when the students' attention was drawn to the fact that the other forces (normal force and weight) were the same in the free body diagrams of both problems and they are both equilibrium problems, only some of the students appeared concerned. Others used convoluted reasoning and asserted that friction has a special formula which should be used whereas tension does not have a formula, and therefore, the free body diagram must only be used for problems involving tension (not problems involving friction).

## 3  Students' Performance on Analogical Reasoning

We then conducted a related study in another algebra-based introductory physics course with 37 students. In a recitation class, students were given the problem with tension as a solved problem (see Appendix I). This time they were explicitly asked to make an analogy between the solved problem involving tension and the quiz problem involving friction and exploit the analogy to solve the quiz problem. We find that, even with an explicit instruction to make an analogy with the problem involving tension, only 35% of the students were able to solve the quiz problem involving friction. Thus, two thirds of the students were unable to make a good analogy between the solved problem and the quiz problem. The most common incorrect response, by 16% of the students, was that the magnitude of the frictional force is $\mu_s N$. Other students with incorrect responses used an expression for the frictional force that involved either $\mu_s$ or $\mu_k$ or both. Moreover, five students drew an analogy between the solved problem and the quiz problem and wrote $F = mg \sin\theta = 7500 N$. But they also incorrectly wrote that the frictional force was $\mu_s N$ or they incorrectly multiplied 7500N (which is the correct magnitude of the frictional force) with the coefficient of static or kinetic friction to compute the magnitude of the frictional force incorrectly.

Three examples of students' work showing such difficulties are featured in Appendix II. The first student in Appendix II incorrectly believes that the magnitude of the frictional force is $\mu_s N$. Therefore, instead of correctly writing that the magnitude of the frictional force is $mg \sin\theta$, the student incorrectly writes that $\mu_s N = mg \sin\theta$ (which the student uses first to calculate the normal force $N$ incorrectly). However, the student gets the correct numerical answer in the end by using $\mu_s N = mg \sin\theta$ (because the frictional force is $mg \sin\theta$). If this student had to calculate the normal force, the student would have obtained an incorrect numerical answer. The second student's work in Appendix II shows that the student first correctly calculates the magnitude of the frictional force but then incorrectly believes that he should multiply

$mg\sin\theta$ by the coefficient of static friction to find the frictional force. The third student's work in Appendix II shows that the student is simultaneously using two equations to calculate the frictional force. One of them is the correct equation while the other is the incorrect equation $F = \mu_s N$ which only applies to the maximum value of the static frictional force. Thus, while this student ended up with the correct numerical value of the frictional force, his calculated value of the normal force is incorrect.

## 4  Summary and Conclusions

We find that a majority of students in an algebra-based introductory physics course could not exploit the deep analogy between the solved problem and the quiz problem to solve the quiz problem even when explicitly asked to do so. The students often made analogies that were superficial. The fact that so many students were unable to use the analogy and take advantage of the solved problem about tension in a rope to tackle the problem involving friction suggests that presenting students with a very well-laid-out analogy with which to solve a difficult problem doesn't always work.

One framework for understanding why an analogy that appears to be promising to an expert doesn't always work emphasizes the fact that introductory physics students often don't value consistency and that they tend to think of the subject as a jumble of disconnected formulas [1-3, 18-19]. Because students have no expectation of deep similarities, they resort to memorized formulas [1-3, 18-19]. For example, many students had the misleading notion that information about frictional force is required and they relied incorrectly on the static friction being equal in magnitude to $\mu_s N$ instead of recognizing the link between paired problems. This theoretical position suggests that the analogy strategy may not work because students don't expect physics to be a coherent body of knowledge which leads them to disregard the rope problem when solving the friction problem.

The instructors must realize that they are able to discern the deep similarities between the solved problem and the quiz problem because of their vast experience and their emphasis on coherence in the physics knowledge. Their robust knowledge structure and their search for coherence facilitates analogical reasoning but it does not imply that their students in the introductory physics courses will be able to do the same without explicit guidance. This expectations-based interpretation suggests that students will be able to exploit analogical reasoning effectively only if instructional strategies embedded within a coherent curriculum forces students to learn to value coherence in problem-solving, e.g., by modeling it in class and grading for it on exams.

## 5 Acknowledgments

We thank F. Reif, R. Glaser, R. P. Devaty and J. Levy for helpful discussions.

## Appendix I: Quiz and Solved Problem

You are asked to exploit the similarities between a solved problem provided to you and your quiz problem in order to solve the quiz problem. Before you solve the quiz problem, go over the solved problem carefully and then answer the questions below.

**Quiz Problem:**

**A car which weighs** 15,000 **N is at rest on a** $30^0$ **incline, as shown below. The coefficient of static friction between the car's tires and the road is** 0.90**, and the coefficient of kinetic friction is** 0.80**. Find the magnitude of frictional force on the car.**

Note: These trigonometric results might be useful: sin $30^0$=0.5, cos $30^0$=0.866.

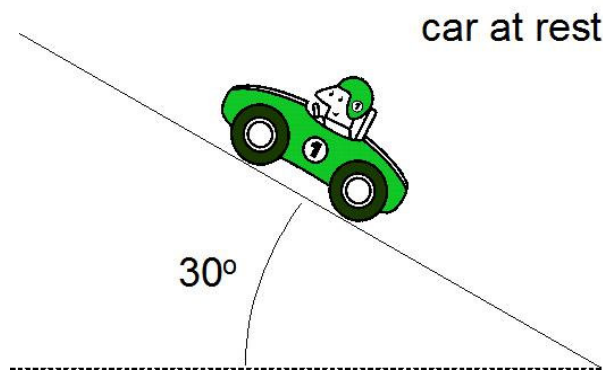

a) Explicitly write down in detail the similarities between the quiz problem and the solved problem provided and how you can exploit the similarities to solve the quiz problem.

b) Now solve for the magnitude of the frictional force on the car in the quiz problem above.

**Solved Problem:**

**A car which weighs 15,000 N is at rest on a frictionless $30^0$ incline, as shown below. The car is held in place by a light strong cable parallel to the incline. Find the magnitude of the tension force in the cable.**

Note: These trigonometric results might be useful: sin $30^0$=0.5, cos $30^0$=0.866.

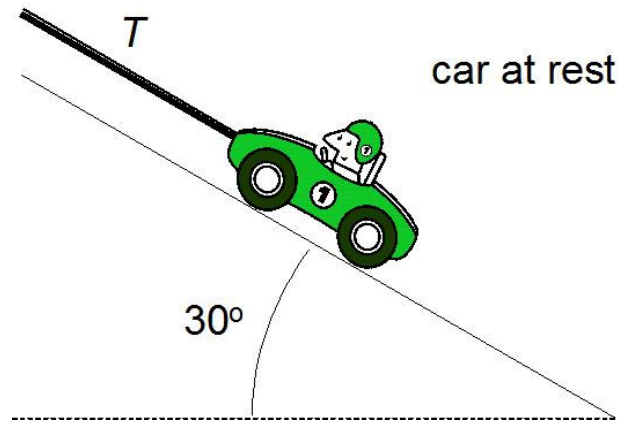

**Solution:**

DESCRIPTION OF THE PROBLEM:

**Knowns:**

magnitude of the weight of the car of mass *m*: W = *mg* = 15000N

angle of the incline: $\theta = 30^0$

The incline is frictionless.

**Target Quantity:**

the magnitude of the tension force (*T*)

CONSTRUCTING THE SOLUTION:

**Free body diagram:**

Since the incline is frictionless, there are only 3 forces acting on the car of mass *m*: the gravitational force (magnitude *mg*), the normal force (magnitude *N*), and the tension (magnitude *T*). Because the car is stationary, the velocity of the car, which is zero, does not change with time. Therefore, the acceleration, $\vec{a}$, of the car is zero, i.e., $\vec{a} = 0$.

From Newton's second Law: $\vec{F}_{net} = m\vec{a}$

$\vec{F}_{net}$ is defined as the vector sum of all the forces acting on the car and should be zero:

$\vec{F}_{net} = \vec{0}$ since $\vec{a} = 0$.

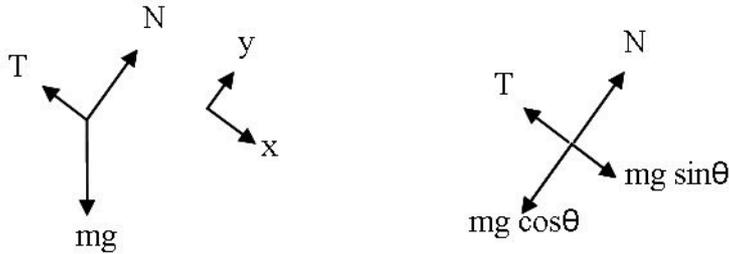

We can decompose the force into x and y components after choosing appropriate coordinate axes as shown.

Both $\vec{F}_{net,x} = \vec{0}$ and $\vec{F}_{net,y} = \vec{0}$.

From $\vec{F}_{net,x} = \vec{0}$ we have $mg \sin\theta - T = 0$

So $T = mg \sin\theta = 15000N \cdot \sin 30^0 = 7500N$

REASONABILITY CHECK OF THE SOLUTION:

**Limiting case 1:** $\theta = 0$, we expect $T$ to be zero, which agrees with our result for $T$:
$T = mg \sin\theta = mg \sin 0 = 0$

**Limiting case 2:** $\theta = 90^0$, we expect the magnitude of tension $T$ to be equal to the weight of the car, which agrees with our result for $T$:
$T = mg \sin\theta = mg \sin 90^0 = mg$

# Appendix II: Three Sample Students' Solutions Showing Misconceptions related to Friction

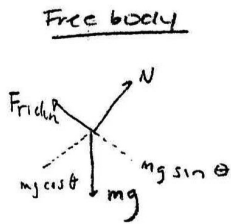

$m = 15,000 N$   angle of incline = $30°$   $\mu_s = .90$

Free body

at rest : $V = 0$   so $a = 0$

$F_s = \mu_s N$

$F_{net} = ma$ ; $F_{net} = 0$

$F_{net\,x} = 0$

$F_s - mg\sin\theta = 0$

$\mu_s N - mg\sin\theta = 0$

$\mu_s N = mg\sin\theta$

$.9 N = (15000 N)\sin 30$

$N = \dfrac{(15000)(.5)}{.9}$

$N = 8333.33$

$F_s = \mu_s N$

$F_s = .9 (8333.33)$

$\boxed{F_s = 7500 N}$

$F_k - mg \sin \theta = 0$

$\therefore F_k = mg \sin \theta$

$\phantom{\therefore F_k} = 15,000 \text{ N} \cdot \sin 30°$

~~$F_k = 7500 (N) \cdot .90 =$~~    $6750 \text{ N} = F_s$

$\phantom{F_k = 7500 (N) \cdot .90 = } 6000 \text{ N} = F_k$

$\boxed{\Sigma F = 12,750 \text{ N}}$

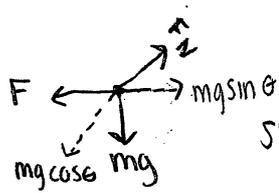

$\mu_s = .90 \quad \mu_k = .80$

since $a = 0$, $F_{net} = 0$

$F_{net\,x} = 0$
$F_{net\,y} = 0$

$F_{net}\,X: \quad F - mg\sin\theta = 0$

$F = mg\sin\theta$

$F = 15{,}000\,N \sin 30$

$F = 7500\,N$

$W = mg$
$= 15000\,N$

$\theta = 30°$

$F = \mu_s |F_N|$
$7500 = .90\,|F_N|$
$F_N = 8333\,N$